\documentstyle[preprint,prb,aps,epsf,floats]{revtex}
\tightenlines
\begin{document}


\title{Cubic boron nitride: an experimental and theoretical ELNES study}

\author{D.N. Jayawardane \and Chris J. Pickard \and L.M. Brown \and M.C. Payne}
\address{Cavendish Laboratory, University of Cambridge, Madingley
road, Cambridge CB3 0HE, UK.}
\date{\today}
\maketitle

\begin{abstract}
A comparison between experimental and theoretical electron Energy Loss
Near Edge Structure (ELNES) of B and N K-edges in cubic boron nitride
is presented.  The electron energy loss spectra of cubic boron nitride
particles were measured using a scanning transmission electron
microscope.  The theoretical calculation of the ELNES was performed
within the framework of density functional theory including single
particle core-hole effects.  It is found that experimental and
calculated ELNES of both the B and N K-edges in cubic boron nitride
show excellent agreement.
\end{abstract}
\pacs{}


\newcommand{\ket}    [1]{{|#1\rangle}}
\newcommand{\bra}    [1]{{\langle#1|}}
\newcommand{\braket} [2]{{\langle#1|#2\rangle}}
\newcommand{\bracket}[3]{{\langle#1|#2|#3\rangle}}

\section{Introduction}

There has been an increasing technological interest in cubic boron
nitride (c-BN).\cite{pouch90,edgar92,vel91} Among the super hard
materials, c-BN has particularly fascinating properties, such as high
thermal stability, low chemical reactivity, especially with ferrous
materials, and an ability to be doped both n or p type. It can thus
compete with diamond in many potential
applications.\cite{edgar92,vel91,mishima90} Like diamond, c-BN is an
sp$^3$ bonded structure and exhibits a stacking sequence of
AaBbCcAa\dots in the [111] direction.  However, it is difficult to
synthesise pure c-BN and most cases it either comes as a mixture of
phases\cite{mckenzie91} or crystallises as a variable composition
compound with defects and stacking faults.\cite{bezhenar98}

Electron Energy Loss Spectroscopy (EELS) is a premier diagnostic
technique, especially for sub micron particles, since the focussing of
the electron probe allows measurements to be taken from extremely
small samples. In particular, there is great interest in the Energy
Loss Near Edge Structure (ELNES). It can yield information on the
bonding, local electronic structure and nearest neighbour coordination
of the excited atom.\cite{egerton96,rez92} This so-called coordination
fingerprint can be used for the identification and quantification of
unknown phases in complex systems such as the borides, nitrides,
carbides and so on.\cite{brydson92,hofer87} The ELNES spectra of a
sample can be obtained with a parallel EELS system mounted on a
Scanning Transmission Electron Microscope (STEM) equipped with field
emission gun (FEG).\cite{mcmullan92} In our case the FEG has an energy
resolution of about 0.3V. Such a system can be competitive with X-ray
Absorption Spectroscopy (XAS), especially where only small samples are
available, and for the lighter elements.

The ELNES is the result of the transition of tightly bound core
electrons to the unoccupied conduction band states induced by the
passing of a swift electron. Hence ELNES probes the local density of
unoccupied electronic states.  Calculations of ELNES using
self-consistent band theory methods\cite{rez91,kostlmeier99} and
multiple scattering theory\cite{rehr92,merchant98,wibbelt99} show
reasonable overall agreement with available experimental
spectra. Specifically, the ELNES of c-BN has been presented by many
authors, both theoretically\cite{rez91,merchant98,wibbelt99} and
experimentally.\cite{schmid95,jaouen95} However, the detailed
agreement between the experimental and calculated c-BN ELNES spectra
has not been perfectly satisfactory. Either the resolution of the
experimental spectra has been insufficient to reveal all the fine
structure features produced in the theoretical calculations or the
theoretical spectra have not fully reproduced the experimental fine
structure. Although the correct number of observed peaks may be given
by theory, their relative positions and intensities are often not
correctly predicted.  In this paper we present a comparison between
state-of-the-art experimental (see Section \ref{exp}) and calculated
(see Section \ref{calc}) c-BN ELNES. Not only do the new experimental
results reveal practically all of the fine features found in the
calculations, the positions and intensities of these features agree up
to 60eV above the threshold.

\section{Experiment}
\label{exp}

The c-BN powder used in these studies was produced and supplied to us
by de Beers. These microcrystallites, a few microns in size and white
in colour, were grown by a high-pressure and high-temperature route
using hexagonal boron nitride (h-BN) as the starting material.

The specimens were prepared for the STEM measurements by grinding some
of the powder using an agate pestle and mortar and then mixing it with
a few drops of water. The tiny particles dispersed in the water were
then collected onto a holey carbon film.  The specimen was baked for a
few hours in the microscope or exposed to the electron beam for some
time to eliminate or reduce any carbon or other contamination.  The
EELS measurements were carried out in a Cambridge VG HB501 dedicated
STEM fitted with an improved Parallel Electron Energy Loss
Spectroscopy (PEELS) system by McMullan.\cite{mcmullan92} The HB501 is
equipped with cold field emission gun operating at 100KV and has an
energy spread of about 0.25eV.  At present, our PEELS system gives a
spectral resolution of about 0.35eV, as given by the full width at the
half maximum of the zero-loss peak. Absolute energy measurements can
be obtained to an accuracy of $\pm$0.3eV, as judged by the $\pi^*$
peak energy of C K-edge in graphite. We find it to be
285.37$\pm$0.3eV, compared to a reported absolute energy
value\cite{batson93} of 285.38$\pm$0.05eV. Sufficiently thin regions
of a c-BN flake suspended on a hole in the carbon support film were
selected for recording spectra.  All the spectra were acquired under
the same conditions of a beam convergence semiangle of 13.2 mrad, and
an effective spectrometer collection semiangle of 13 mrad.  Under
these conditions, to a good approximation, non-dipole transitions are
not allowed because, for light elements, the inelastic scattering is
strongly forward peaked.\cite{leapman81} For the ELNES region, spectra
were acquired with an energy dispersion of 0.23eV/channel, in order to
include the full range of the ELNES in one spectral region. Also,
energy dispersions of 0.1eV/channel and 0.8eV/channel were used to
acquire the spectra, to analyse the band gap region and the
stoichiometry (B/N) respectively in the same region of the BN
particle.  In each case, a minimum of 10 spectra with a high number of
counts each were acquired using long exposure times.

The processing of the spectra was accomplished using PEELS software
developed in house.  First, the dark-current was subtracted from all
the spectra and flat fielded for a channel-to-channel gain correction.
Second, each set of spectra (a minimum of 10) was aligned using a
least square fitting algorithm and summed.  From the resulting high
loss region of the spectra the background was subtracted using the power
law form, AE$^{-r}$. Then, to remove the effects of multiple
scattering, the spectra were deconvolved by the raw low loss spectra (a
Fourier-ratio deconvolution).\cite{egerton96} Finally, each resulting
spectrum was deconvoluted by the zero-loss spectrum to remove the
point-spread function.

\section{Calculation}
\label{calc}

The electron energy loss function, for high-energy losses, is directly
proportional to the imaginary part of the dielectric function,
$\varepsilon_2({\bf q},E)$.  Hence, within the single particle
approximation, the measured intensity distribution of ELNES on the K-edge
for a periodic system is given by:
\begin{equation}
\varepsilon_2({\bf q},E) = \frac{2\pi e^2}{\Omega\varepsilon_0
q^2}\sum_{{\bf k},{\rm c.b.}}|\bracket{\psi_{\bf k}^{\rm
c.b.}}{e^{i{\bf q.r}}}{\psi_{1s}}|^2\delta(E_{\bf k}^{\rm
c.b.}-E_{1s}-E)
\end{equation}
where $\Omega$ is the unit cell volume, ${\bf q}$ is the momentum
transfer and ${\bf k}$ is the k-point within the first Brillouin zone
of the final state in the conduction band ($\bra{\psi_{\bf k}^{\rm
c.b.}}$). The K-edge is given by the excitation from a $1s$ core state
($\ket{\psi_{1s}}$), so this is the initial state in the above
expression. The energies $E_{1s}$ and $E_{\bf k}^{\rm c.b.}$ denote
the energy levels of the initial and final states respectively.

For the calculation of ELNES, the two main tasks are the evaluation of
the density of unoccupied electronic states (DOS) term and the
weighting transition matrix elements.  In our scheme (which is
described in detail by Pickard\cite{pickard-thesis} and
elsewhere\cite{pickard97,rez99}) the final states and energies
$\ket{\psi_{\bf k}^{\rm c.b.}}$ and $E_{\bf k}^{\rm c.b.}$ are
evaluated within an \emph{ab initio} electronic structure calculation;
using a total energy code (CASTEP) based on Density Functional Theory
(DFT) within the Local Density Approximation (LDA).  A plane wave
basis set, periodic boundary conditions and non-local pseudopotentials
are used.\cite{payne92} Hence, the summation over all final states is
the sum over all ${\bf k}$ within the first Brillouin zone and all
unoccupied bands ${\rm c.b.}$ at each k-point.  An efficient Brillouin
zone integration scheme has been adopted, which uses a very low
k-point sampling and is based on an extrapolative
approach.\cite{pickard99} The information for the extrapolation is
obtained using {\bf k.p} perturbation theory to second order within a
set of sub volumes into which the Brillouin zone is divided.  Some
corrections to the {\bf k.p} expansion are made for the use of
non-local pseudopotentials\cite{pickard2000}.  The resulting piecewise
quadratic representation of the band structure is directly converted
into a DOS using the analytic quadratic approach of
Methfessel.\cite{methfessel83}
	
The representation of the wavefunctions in terms of plane waves allows
the direct quantitative evaluation of the matrix element, in
contrast to the more frequent approaches which are simply symmetry projected
local DOS.  In our evaluation of the matrix elements, the initial
core states are taken from all electron calculations for the isolated
atoms.  In practice, the dipole approximation is applied.  The directly
evaluated matrix elements are then corrected for the pseudopotential
error using the Projector Augmented Wave approach of van de Walle and
Blochl.\cite{vandewalle93} Matrix elements evaluated in this way allow the
absolute prediction of the cross-section with the correct dipole selection
rules.

In order to compare our calculations with measured ELNES, particularly
in insulators, it is important to consider the effects of the
excitation process, in particular single particle core hole
effects.  These derive from the un-screening of the nuclear charge as
an electron is excited from a localised core-state.  In contrast to
the situation in metals, in insulators the effect due to the core hole
can be dominant, as the re-screening of the hole by the valence
electrons is not complete.  In our approach, the effects of the
interaction between the core-hole and the ejected electron are treated
by band structure methods but the remaining many body effects, which
would lead to multiplet structure, are ignored.
	
In our current scheme, the effects of the core-hole on the unoccupied
electronic states for systems with moderate core hole effects (eg. BN,
diamond) are explicitly calculated by performing a supercell
calculation in which there is a single excited potential in each cell.
The supercell must be large enough that the neighbouring excited
potentials do not interact with each other. For the c-BN system, the
calculations are performed in a supercell comprising $2^3$ primitive
lattice cells.  Then the spectra are evaluated for the excited
atom. To model the excited atom, the more usual Z+1 approximation
which would substitute, say, B by C, is improved upon by generating a
special pseudopotential for the excited B atom with only one $1s$
electron.

\section{Results and Discussion}

\begin{figure}
\centerline{
\epsfxsize=7cm\epsfbox{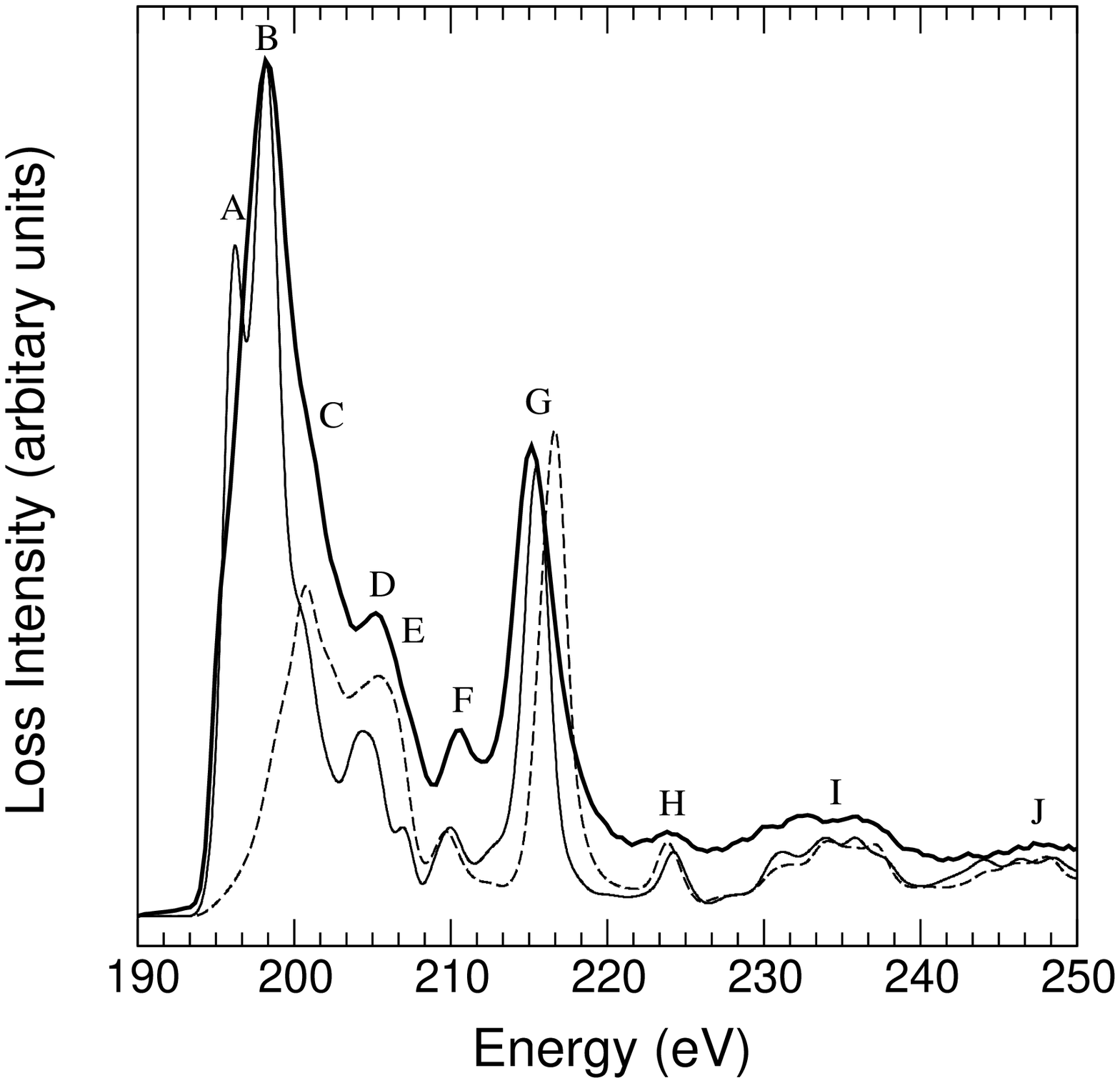}
}
\centerline{
\epsfxsize=7cm\epsfbox{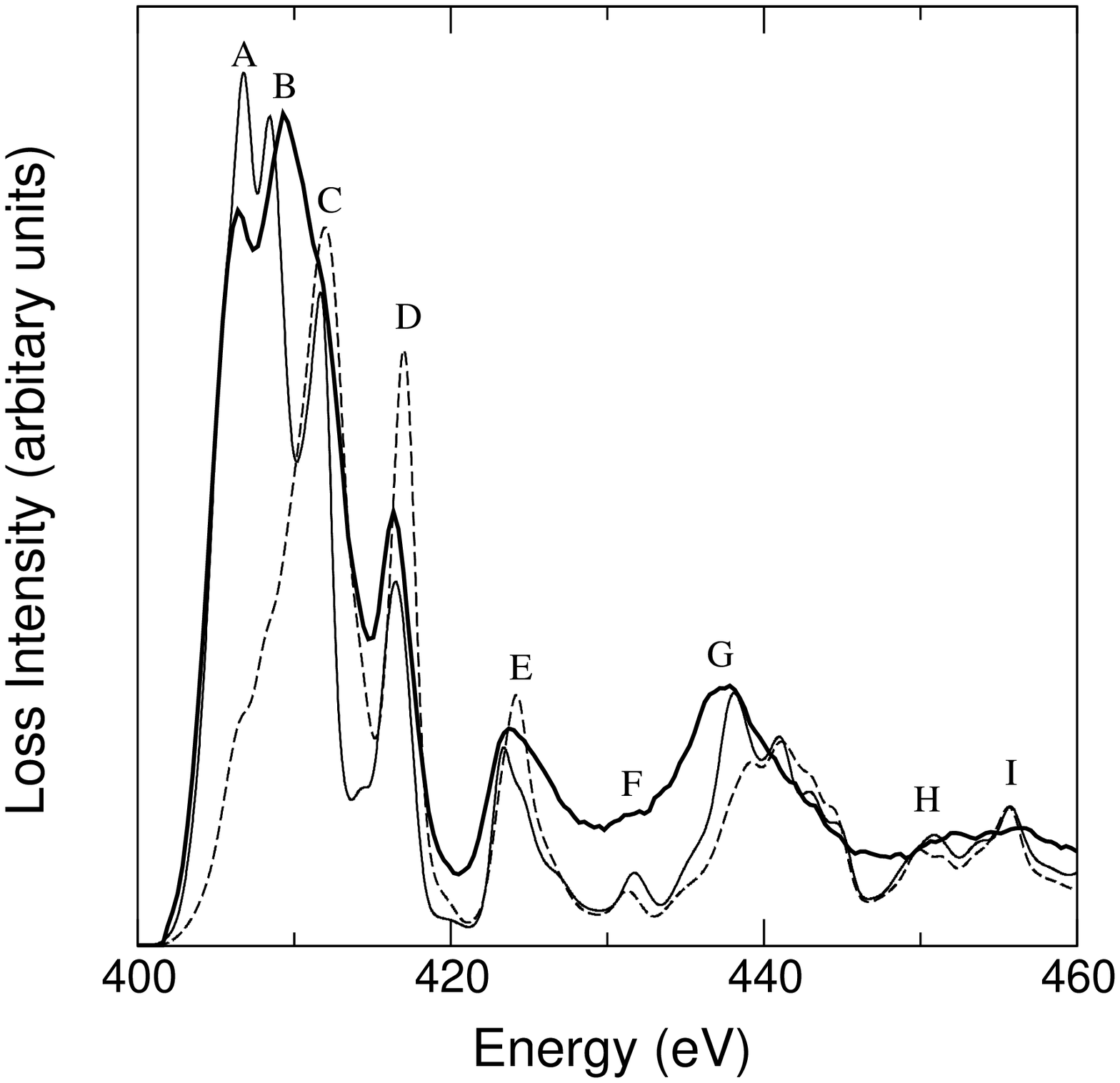}
}
\caption{\label{fig1} Experimental ELNES (thick solid lines),
theoretically calculated ELNES with core-hole effects (thin solid
lines), and theoretically calculated ELNES without core-hole effects
(thin dotted lines) of the boron K-edge (top) and the nitrogen K-edge
(bottom) in cubic boron nitride.}
\end{figure}

The quality of the c-BN sample was assessed by X-ray powder
diffraction (XRD) and EELS.  The XRD pattern of the c-BN powders shows
non-distorted peaks with relative intensity and $d$ spacings very
close to the reported c-BN standard XRD values (JCPDS 35-1365).  The
lattice parameter deduced by XRD is 3.616$\pm$0.002 (the reported
standard values are a = 3.615 or 3.620$\pm$0.001).  The EELS study of
the composition and the stoichiometry of the particles confirm that
the crystals used in our experiments are stoichiometric and free from
contamination.  The ELNES given in Figure \ref{fig1} is from a defect
free region of a c-BN particle about one mean free path thick which
gives stoichiometry B/N=1$\pm$0.03 The bulk plasmon peak is measured
to be 32.4$\pm$0.2eV.

Figure \ref{fig1} shows the comparison between the experimental and
calculated ELNES of the B and N K-edges of c-BN. The figure shows
theoretical calculations with core-hole (thin solid lines) and without
core-hole effects (dotted lines). Each calculated K edge is aligned to
the energy of the peak labelled B and D for B and N K-edges
respectively and the intensity is normalised to the height of peak B
of the experimental spectra. While the calculated spectra is aligned
with the experimental spectra as described, the energy scale is not
adjusted.  In the experimental ELNES, the features present show a
breadth of 4eV or more, much greater than the experimental resolution.

At first glance, one can recognise the excellent agreement between the
experimental and theoretical ELNES of both B and N K-edges. Both the
peak positions and the relative intensities seen in the experimental
results are reproduced by the calculations.  A tabulation of the
energies of the features in the experimental and calculated ELNES is
given in Table \ref{tab1}. The calculated spectra are convolved with a
Gaussian whose width is 0.5eV which we believe to be comparable to the
spectral resolution. However, lifetime broadening effects have been
excluded from our theoretical calculations.

As can be seen by the comparison between the solid and dotted lines in
Figure \ref{fig1}, core-hole effects dominate both the B and N K-edge
ELNES at the threshold.  By including core-hole effects, ELNES
features up to 15eV above the edge onset are modified, giving close
agreement to the experimental ELNES. Peaks A and B are particularly
strongly modified by core hole effects. In the calculated ELNES, peak
A at the edge onset has been enhanced over the experimental features
for both the B and N K-edges.  This may be due to an over estimation
of the strength of the core hole.  Up to 30eV above the threshold, the
experimental ELNES features of both the B and N K-edges of c-BN are
very similar to the equivalent XANES results of Chaiken \emph{et
al}.\cite{chaiken93} Indeed, the absolute energy values are in close
agreement with the XANES values (B K-edge: 194.5, 198.0 eV and N
K-edge: 406.0, 409.0, 411.0 eV for peaks A,B,\dots respectively) given
by Jimenez \emph{et al}.\cite{jimenez97} Peak A of the B K-edge is
sharper and shifted to the lower energy in XANES, by comparison to
that in the ELNES measurement.  As described by Batson,\cite{batson91}
this may be due to a reduction of the excitonic distortion by the
incident swift electron in ELNES.  Furthermore, this theoretical
scheme correctly represents the EXELFS features, (G, H, I) up to 60eV
above the edge, while at the same time, the peaks dominated by
core-hole effects (up to 15eV from the threshold) are adequately
enhanced. This is a feature particular to the plane wave approach, in
that the upper and lower regions of the spectra are equally well
represented by the basis set.

In summary, our experimental and theoretical ELNES show excellent
agreement provided that core-hole effects are included. However, we
expect that further improvements will be made at higher energies by
including lifetime broadening effects,\cite{kostner-thesis} and at the
threshold by reducing the strength of the core-hole interaction.

\section*{Acknowledgement}

DNJ would like to thank the de Beers Industrial Diamond Division for the
specimen and financial support. She would also like to thank New Hall College
and The Cambridge Commonwealth Trust. CJP acknowledges the support of
an EPSRC research studentship.



%
%

%
%

\begin{table}
\caption{Comparison of the energies of the features in the experimental and calculated ELNES for the boron and nitrogen K-edges. The asterix (*) denotes the alignment peak in each case.}
\label{tab1}
\begin{tabular}{lccccccccc}
Boron &    A  &    B$^*$  &    C  &    D  &    E  &    F  &    G  &    H  &   \\
Exp.  & 195.5 & 198.1 & 201.0 & 205.3 & 207.2 & 210.6 & 215.2 & 224.0 &   \\
Calc. & 196.2 & 198.1 & 200.3 & 204.4 & 207.0 & 209.9 & 215.4 & 224.2 &   \\\hline
Nitrogen &    A  &    B  &    C  &    D$^*$  &    E  &    F  &    G  &    H  & I \\
Exp.     & 406.4 & 409.4 & 411.5 & 416.4 & 423.7 & 431.2 & 437.5 &       &   \\
Calc.    & 406.7 & 408.4 & 411.6 & 416.4 & 423.3 & 431.5 & 437.9 & 450.8 & 455.7 \\
\end{tabular}
\end{table}


\begin{thebibliography}{10}

\bibitem{pouch90}
 in {\em Synthesis and properties of boron nitride}, edited by J. Pouch and S.
  Alterovitz (Trans-Tech, Switzerland, 1990).

\bibitem{edgar92}
J. Edgar, J. Mater. Res. {\bf 7},  235  (1992).

\bibitem{vel91}
L. Vel, G. Demazeau, and J. Etourneau, Mater. Sci. Eng. B, Solid State Mater.
  Ad.Technol. {\bf 10},  149  (1991).

\bibitem{mishima90}
O. Mishima,  in {\em Synthesis and properties of boron nitride}, edited by J.
  Pouch and S. Alterovitz (Trans-Tech, Switzerland, 1990), pp.\ 313--328.

\bibitem{mckenzie91}
D. McKenzie {\it et~al.}, J. Appl. Phys. {\bf 70},  3007  (1991).

\bibitem{bezhenar98}
N. Bezhenar, Journal of super hard materials {\bf 20},  11  (1998), and
  references therein.

\bibitem{egerton96}
R. Egerton, {\em Electron Energy Loss Spectroscopy in the Electron Microscope}
  (Plenum, New York, 1996).

\bibitem{rez92}
P. Rez,  in {\em Transmission Electron Energy Loss Spectroscopy in Materials
  Science}, {\em EMPMD Monograph Series}, edited by M. Disko, C. Ahn, and B.
  Fultz (The Minerals, Metals \& Materials Society, Warrendale, Pennsylvania,
  1992), Chap.~5, p.\ 110.

\bibitem{brydson92}
R. Brydson, H. Sauer, and W. Engel,  in {\em Transmission Electron Energy Loss
  Spectroscopy in Materials Science}, {\em EMPMD Monograph Series}, edited by
  M. Disko, C. Ahn, and B. Fultz (The Minerals, Metals \& Materials Society,
  Warrendale, Pennsylvania, 1992).

\bibitem{hofer87}
F. Hofer and P. Golob, Ultramicroscopy {\bf 21},  379  (1987).

\bibitem{mcmullan92}
D. McMullan, P. Fallon, Y. Ito, and A. McGibbon,  in {\em Electron Microscopy
  (EUREM) 92}, edited by A.~R. \emph{et al} (Universidad de Granada, Granada,
  Spain, 1992), Vol.~1, pp.\ 103--104.

\bibitem{rez91}
P. Rez, X. Weng, and H. Ma, Micros. Microanal. Microstruct. {\bf 2},  143
  (1991).

\bibitem{kostlmeier99}
S. Kostlmeier, C. Elsasser, and B. Mayer, Ultramicroscopy {\bf 80},  145
  (1999).

\bibitem{rehr92}
J. Rehr, R. Albers, and S. Zabinsky, Phys. Rev. Lett. {\bf 69},  3397  (1992).

\bibitem{merchant98}
A. Merchant, D. McCulloch, and R. Brydson, Diamond and Related Materials {\bf
  7},  1303  (1998).

\bibitem{wibbelt99}
M. Wibbelt, H. Kohl, and P. Kohler-Redlich, Phys. Rev. B {\bf 59},  11739
  (1999).

\bibitem{schmid95}
H. Schmid, Micros. Microanal. Microstruct. {\bf 6},  99  (1995).

\bibitem{jaouen95}
M. Jaouen {\it et~al.}, Micros. Microanal. Microstruct. {\bf 6},  127  (1995).

\bibitem{batson93}
P. Batson, Phys. Rev. B {\bf 48},  2608  (1993).

\bibitem{leapman81}
R. Leapman, L. Grunes, P. Fejes, and J. Silcox,  in {\em EXAFS Spectroscopy},
  edited by B. Teo and D. Joy (Plenum, New York, 1981), p.\ 217.

\bibitem{pickard-thesis}
C. Pickard, Ph.D. thesis, Cambridge University, Cavendish Laboratory, 1997.

\bibitem{pickard97}
C. Pickard and M. Payne, Inst. Phys. Conf. Ser {\bf 153},  179  (1997).

\bibitem{rez99}
P. Rez, J.~R. Alvarez, and C. Pickard, Ultramicroscopy {\bf 78},  175  (1999).

\bibitem{payne92}
M. Payne {\it et~al.}, Reviews of Modern Physics {\bf 64},  1045  (1992).

\bibitem{pickard99}
C. Pickard and M. Payne, Physical Review B {\bf 59},  4685  (1999).

\bibitem{pickard2000}
C. Pickard and M. Payne, Physical Review B {\bf 62},  4383  (2000).

\bibitem{methfessel83}
M. Methfessel, M. Boon, and F. Mueller, J. Phys. C: Solid State Phys. {\bf 16},
   l949  (1983).

\bibitem{vandewalle93}
C.~V. de~Walle and P. Bl\"{o}chl, Physical Review B {\bf 47},  4244  (1993).

\bibitem{chaiken93}
A. Chaiken {\it et~al.}, Appl. Phys. Lett. {\bf 63},  2112  (1993).

\bibitem{jimenez97}
I. Jimenez {\it et~al.}, Phys. Rev. B {\bf 55},  12025  (1997).

\bibitem{batson91}
P. Batson and J. Bruley, Phys. Rev. Lett. {\bf 67},  350  (1991).

\bibitem{kostner-thesis}
M. Kostner, Master's thesis, Vienna University of Technology, 2001.

\end{thebibliography}
\end{document}